\documentclass[epj,final]{svjour}

\usepackage{graphicx}
\usepackage{amsmath}
\usepackage{amssymb}
\usepackage{bm}
\usepackage{color}

\usepackage{ae}

\renewcommand{\vec}[1]{{\mathbf{#1}}}

\newcommand{\fref}[1]{Fig.~\ref{#1}}
\newcommand{\Fref}[1]{Figure~\ref{#1}}
\newcommand{\eref}[1]{Eq.~\eqref{#1}}
\newcommand{\Eref}[1]{Equation~\eqref{#1}}

\newcommand{\D}{\ensuremath{\text{d}}}
\newcommand{\E}{\ensuremath{\operatorname{e}}}
\newcommand{\kb}{\ensuremath{k_\text{B}}}
\newcommand{\calH}{\ensuremath{{\cal H}}}
\newcommand{\rousecorr}{\ensuremath{C}}

\newcommand{\blength}{\ensuremath{l}}

\newcommand{\ke}{\ensuremath{c_\text{e}}}
\newcommand{\kp}{\ensuremath{c_\text{P}}}
\newcommand{\kone}{\ensuremath{c_1}}
\newcommand{\be}{\ensuremath{b_\text{e}}}
\newcommand{\Ree}{\ensuremath{R_\text{e}}}

\newcommand{\sblob}{\ensuremath{g}}

\begin{document}


\title{Static Rouse Modes and Related Quantities:\\ Corrections to Chain Ideality in Polymer Melts}
\authorrunning{H. Meyer {\em et al.}}
\titlerunning{Static Rouse Modes and Corrections to Chain Ideality in Polymer Melts} 
\author{H. Meyer, J. P. Wittmer, T. Kreer, P. Beckrich, A. Johner, J. Farago, and J. Baschnagel}
\institute{Institut Charles Sadron, CNRS, 23 rue du Loess, 67037 Strasbourg Cedex, France}
\date{Draft version: \today}

\abstract{Following the Flory ideality hypothesis intrachain and interchain excluded volume interactions are supposed to compensate each other in dense polymer systems.  Multi-chain effects should thus be neglected and polymer conformations may be understood from simple phantom chain models.  Here we provide evidence against this phantom chain, mean-field picture.  We analyze numerically and theoretically the static correlation function of the Rouse modes.  Our numerical results are obtained from computer simulations of two coarse-grained polymer models for which the strength of the monomer repulsion can be varied, from full excluded volume (`hard monomers') to no excluded volume (`phantom chains').  For nonvanishing excluded volume we find the simulated correlation function of the Rouse modes to deviate markedly from the predictions of phantom chain models.  This demonstrates that there are nonnegligible correlations along the chains in a melt. These correlations can be taken into account by perturbation theory.  Our simulation results are in good agreement with these new theoretical predictions.         
} 

\maketitle


\section{Introduction}
In 1953 P.E. Rouse \cite{rouse1953} proposed a model to describe the dynamics of a polymer chain, which has become an important concept in polymer physics \cite{DoiEdwards,RubinsteinColby}.  The Rouse model assumes the chain to be a sequence of Brownian beads.  The beads are connected by entropic springs and immersed in a structureless medium which exerts on every bead two forces, a local random force and a local friction force.  Both forces are linked by the fluctuation-dissipation theorem to ensure correct equilibrium properties \cite{DoiEdwards}. 

This bead-spring model thus considers only local interactions.  Nonlocal interactions, such as hydrodynamic or excluded-volume forces between distant beads along the chain, are ignored.  While this assumption is certainly not appropriate for dilute solution in good solvents, it may be valid in concentrated solutions or polymer melts, where both nonlocal interactions are expected to be screened \cite{DoiEdwards,RubinsteinColby}.  In particular in polymer melts, the screening is supposed to extend down to the monomer level, implying that polymer conformations correspond to those of ideal random walks.  Hence, it is generally believed that the Rouse theory provides, at long times and large length scales, a viable description of the conformational dynamics of polymer melts if entanglements with other chains, giving rise to reptation motion \cite{DoiEdwards,RubinsteinColby,McLeish_AdvPhys2002}, are not important.  

Experimental or computational scrutiny of the Rouse model has thus focused on the behavior of short chains in a melt (for review see e.g.\ \cite{McLeish_AdvPhys2002,PaulSmith_RPP2004}).  These tests reveal that the Rouse model represents a good approximation, but quantitative agreement is hard to obtain.  To explain the observed deviations there are, roughly speaking, two main ideas in the literature: additional intrachain contributions not accounted for by the theory (chain stiffness \cite{HarnauEtal:EPL1999,KrushevEtal_Macromolecules2002,BulacuGiessen:JCP2005}, local excluded volume effects \cite{KreerBaschnagel2001,MolinEtal:JPCM2006}, torsional transitions \cite{RichterMonkenbuschAllgeier1999,ArbeEtal:Macro2001,AllegraGanazzoli:Review1989}) and multi-chain effects which invalidate the phantom-chain-in-a-structureless-medium approach of the Rouse model \cite{PaulSmith_RPP2004}.

In the present article, we provide further evidence for the latter point of view.  However, there is one important respect in which the objective of our discussion is limited.  The Rouse model is, in the first place, an attempt to describe dynamic features of polymer melts.  Here we shall not be concerned with dynamics, but rather study, by theory and simulation, the initial (static) value of the correlation function of the Rouse modes.  This initial value informs us about conformational properties of a chain in the melt, and clearly, an understanding of these equilibrium features is a prerequisite for an extension of the discussion to polymer dynamics.  Our key point is that the conformational properties, even for very flexible chains, are more complex than commonly assumed because chains are not phantom chains, free of any interaction.  In the melt, polymer segments of size $s$ ($s \gg 1$) experience an effective repulsion resulting from chain connectivity and incompressibility of the melt.  This repulsion entails systematic deviations from `ideal' random-walk-like conformations and thus leads to a violation of the `Flory ideality hypothesis' \cite{flory2}, i.e.\ of a central concept of modern polymer physics.  The consequences of these deviations have recently been explored in real (intrachain correlations \cite{WittmerEtal:PRL2004,WittmerEtal:PRE2007}) and reciprocal space (form factor \cite{WittmerEtal:EPL2007,BeckrichEtal:Macro2007}, collective structure factor \cite{SemenovObukhov:JPCM2005}).  The purpose of the present work is to extend this discussion to the static Rouse modes. 

The outline of the article is as follows.  We begin by presenting the simulation models and techniques used in this study (Section~\ref{sec:models}).  Section~\ref{sec:rouse} compares predictions from phantom chain calculations to simulation results of polymer melts with full excluded volume interaction between the monomers.  The comparison reveals systematic deviations between theory and simulation for the static correlation function of the Rouse modes.  These deviations can be rationalized by a theory which accounts for corrections to chain ideality in polymer melts (Section~\ref{sec:fullexvol}).  Section~\ref{sec:finiteoverlap} extends the previous discussion to the situation where the phantom chain limit is approached by gradually making the monomer-monomer interaction softer.  Also this case can be understood by taking corrections to chain ideality into account.  The final section (Section~\ref{sec:summary}) presents a brief synopsis of our results.

\section{Simulation models and techniques}
\label{sec:models}
By computer simulations we examine polymer melts of two coarse-grained models \cite{BaschnagelWittmerMeyer:NIC_Review2004}, a bead-spring model and the bond fluctuation model.  For both models we study a version which precludes monomer overlap (`hard monomers') and another one with finite energy penalty for monomer overlap (`soft monomers').  The latter version allows us to switch gradually from full monomer excluded volume interactions, the standard choice in simulations, to phantom chain behavior, the situation considered by the Rouse model. 

In the following we briefly describe the simulation models and techniques. Further details may be found in Refs. \cite{WittmerEtal:PRE2007,WittmerEtal:preprint2007}.      

\subsection{Bead-spring model}

\subsubsection{Excluded volume chains}  
The bead-spring model (BSM) is derived from a model employed in simulations of polymer crystallization \cite{MeMu01,MeMu02,VettorelMeyer:JCTC2006,VettorelEtal:PRE2007}.  It is characterized by two potentials: a harmonic bond potential,
\begin{equation}
U_\text{b}(r) = \frac{1}{2}k_\text{b} (r - \blength_\text{b})^2 \;,
\label{eq:U_bond}
\end{equation}
where $k_\text{b} = 535.46 \, \kb T / \sigma_0^2$ and $\blength_\text{b} = 0.97234\, \sigma_0$, and a nonbonded 9--6 Lennard-Jones (LJ) potential
\begin{equation}
U_\text{nb}(r) = 
\left\{
\begin{matrix}
\varepsilon_0 & \Big [\left(\frac{\sigma_0}{r}\right)^9 -
\left(\frac{\sigma_0}{r}\right)^6 \Big] + C & & r \le r_\text{min} \;,\\
0 & & & r > r_\text{min} \;, \\
\end{matrix}
\right.
\label{eq:U_mol}
\end{equation}
with $\varepsilon_0 = 1.511\kb T$.  The LJ potential is truncated at its minimum, $r_\text{min} = (3/2)^{1/3} \, \sigma_0$, and shifted to zero by $C = 4 \varepsilon_0/27$.  Intrachain and interchain interactions of nonbonded monomers are thus purely repulsive.  The parameters of the bond potential are adjusted such that the average bond length $\blength \approx \blength_\text{b} = 0.97234\, \sigma_0$ at the monomer density $\rho = 0.84 \, \sigma_0^{-3}$ is very close to that of the standard Kremer-Grest model \cite{KremerGrest1990,AuhlEtal:2003}.  In the following, we report all data in reduced units, that is, energies are measured in units of $\kb T$ (Boltzmann constant $\kb \equiv 1 $) and lengths in units of the monomer diameter $\sigma_0$.

\subsubsection{From hard to soft monomers}
\label{subsubsec:hard2soft}

\begin{figure}
\begin{center}
\includegraphics*[width=0.75\linewidth]{Unb_force_capping_sketch}\\[5mm]
\includegraphics*[width=0.6\linewidth]{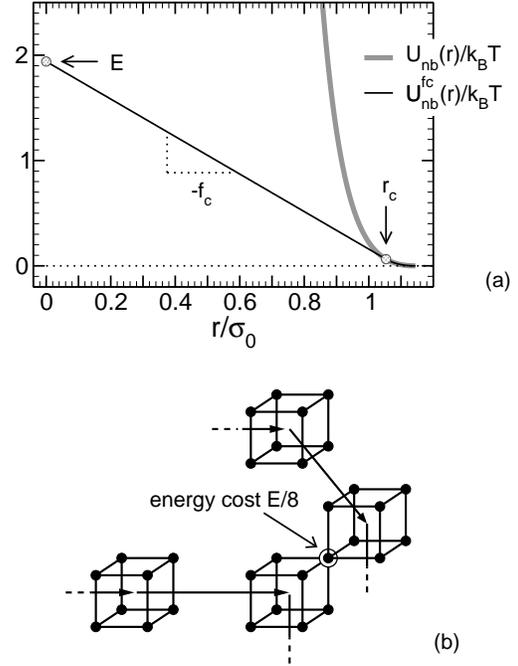}
\end{center}
\caption{Illustration of the modeling of soft monomers for the BSM [panel (a)] and the BFM [panel (b)].  For the BSM a force capping with parameter $f_\text{c}=1.781$ is shown.  This gives $r_\text{c} = 1.0535$ and an energy penalty for monomer overlap of $E=U_\text{nb}(r_\text{c}) + f_\text{c} r_\text{c} = 1.94$.  For the BFM every doubly occupied lattice site is penalized by an energy cost of $E/8$ so that the energy penalty for full overlap of two monomers equals $E$.} 
\label{fig:sketch_softmonomers}
\end{figure}

Equations~\eqref{eq:U_bond} and \eqref{eq:U_mol} ensure that chains cannot cross each other and monomers are impenetrable (`hard monomers'). We `soften' these constraints by introducing a force capping through the following modification of the nonbonded interactions
\begin{equation}
U_\text{nb}^\text{fc}(r) = 
\left\{ 
\begin{array}{ll}
U_\text{nb}(r_\text{c}) + f_\text{c} (r_\text{c} - r) 
                        & \quad 0\le r \le r_\text{c} \;,\\
U_\text{nb}(r)          & \quad r > r_\text{c} \; .
\end{array}
\right.
\label{eq:U_mol_fc}
\end{equation}
Here the `capping' distance $r_\text{c}$ is defined by the condition that the nonbonded force at $r_\text{c}$ has the prescribed value $f_\text{c}$, i.e.\ $-\D U_\text{nb}(r)/\D r |_{r=r_\text{c}} =f_\text{c}$.  This definition implies that $r_\text{c}$ tends to 0 for $f_\text{c} \gg 1$ and to $r_\text{min}$ for $f_\text{c} \rightarrow 0$.  (See \fref{fig:sketch_softmonomers}(a) for an illustration.)  For distances smaller than $r_\text{c}$ \eref{eq:U_mol_fc} thus replaces the steep rise of $U_\text{nb}(r)$ by a much weaker linear increase, entailing a finite energy penalty  $E=U_\text{nb}(r_\text{c}) + f_\text{c} r_\text{c}$ for two overlapping monomers ($r=0$).  The smaller $f_\text{c}$, the softer the monomers.  In particular, if $f_\text{c}=0$, we recover phantom chain behavior, i.e.\ $U_\text{nb}^\text{fc}(r) =0$, because $r_\text{c}$ then equals $r_\text{min}$ and $U_\text{nb}(r)$ vanishes for $r \geq r_\text{min}$.  

In this way, we also simulate two phantom chain models, a (generalized) freely-jointed chain (FJC) model characterized only by the bond potential of \eref{eq:U_bond}, and a (generalized) freely-rotating chain (FRC) model which, in addition to \eref{eq:U_bond}, also presents a potential for the bond angle $\theta$, 
\begin{equation}
U_\text{ang}(\theta) = k_\theta \big [1 - \cos 2 (\theta - \theta_0) \big ] \;,
\label{eq:U_angle}
\end{equation}
with $k_\theta = 10$ and $\theta_0 = 109^\circ$.  For this potential we find that the average cosine of the bond angle is $\alpha = \langle \cos \theta \rangle = -0.320713 \approx \cos \theta_0 $.

\subsubsection{Simulation aspects}
With these models we perform molecular dynamics (MD) simulations at constant temperature $T=1$ (Langevin thermostat with friction constant $\gamma = 0.5$) and constant monomer density $\rho = 0.84$, the typical melt density of the Kremer-Grest model \cite{KremerGrest1990,AuhlEtal:2003}.  The equations of motion are integrated by the Velocity-Verlet algorithm \cite{AllenTildesley}.  For `hard' monomers we combine the MD with double-bridging Monte Carlo (MC) moves \cite{BaschnagelWittmerMeyer:NIC_Review2004,AuhlEtal:2003} to speed up the decorrelation of large-scale conformational features.  As only few of these MC moves are accepted per unit time, this does not deteriorate the stability or accuracy of the MD.  The MC moves, however, considerably improve the statistics for large chain lengths.  For `soft' monomers only MD is used.  We will discuss data for chain lengths (number of monomers per chain) $N=64$, 256, 512, and 1024, obtained from simulations of periodic systems of linear size $L \leq 62$.  For $\rho=0.84$ these systems contain up to $196\,608$ monomers.

\subsection{Bond fluctuation model}

\subsubsection{Excluded volume chains}
We also examine the three-dimensional bond fluctuation model (BFM) on a cubic lattice \cite{DeutschBinder:JCP1991,PaulEtal:JPII1991}.  Each monomer occupies a cube of eight adjacent sites, the length of the bonds between connected monomers along a chain are allowed to fluctuate in the range from 2 to $\sqrt{10}$ lattice constants (the lattice constant will be the length unit in the following), and double occupancy of lattice sites is forbidden by a hard-core interaction between monomers.  The system is athermal, the only control parameter being the monomer density $\rho$.  Melt conditions are realized for $\rho = 0.5/8$, where half of the lattice sites are occupied \cite{PaulEtal:JPII1991}.  We use periodic simulation boxes of linear dimension $L=256$ which contain $\rho L^3 \approx 10^6$ monomers.  These large systems eliminate finite-size effects, even for the longest chain lengths studied ($256 \leq N \leq 8192$).  The simulations are carried out by a mixture of local, slithering-snake, and double-bridging MC moves which allow us to equilibrate polymer melts with chain lengths up to $N=8192$ \cite{WittmerEtal:PRE2007}.  

\subsubsection{From hard to soft monomers}
In analogy to the BSM we also study soft monomers by introducing a finite energy penalty $E/8$ for a doubly occupied lattice site (see \fref{fig:sketch_softmonomers}).  This implies that full overlap between two monomers leads to an energy cost of $E$ (as for the BSM, cf.\ Section~\ref{subsubsec:hard2soft}).  A local or slithering-snake move, leading to $N_\text{ov}$ double-occupancies, gives rise to a total energy $\calH / \kb T = N_\text{ov} E/8$.  With the energies of the final ($\calH_\text{f}$) and initial configurations ($\calH_\text{i}$) we accept the move according to the Metropolis criterion \cite{BaschnagelWittmerMeyer:NIC_Review2004,LandauBinder} with probability $\min ( 1, \exp[-(\calH_\text{f} - \calH_\text{i})/ \kb T])$.

\section{Rouse model: predictions and observed deviations}
\label{sec:rouse}
The basic variables of the Rouse model are the Rouse modes $\vec{X}_p$.  Here we introduce two definitions for $\vec{X}_p$, depending on whether we consider a discrete (simulation) model or a continuous (theoretical) model.

For the discrete case let $\vec{r}_n$ be the position of monomer $n$ ($n=1,\ldots,N$).  The Rouse modes are defined by \cite{Verdier1966} 
\begin{equation}
\vec{X}_p = \frac{1}{N}\sum_{n=1}^N \vec{r}_n  \cos\frac {(n-1/2)p\pi}{N}
\label{eq:rousemodedis}
\end{equation}
with $p=0,\ldots,N-1$.  In the continuum limit we will use the notation of Ref.~\cite{DoiEdwards} and write
\begin{equation}
\vec{X}_p = \frac{1}{N}\int_{0}^N \D n\, \vec{r}(n)  \cos\frac {np\pi}{N}
\quad (p=0,1,2,\ldots) \; .
\label{eq:rousemodecon}
\end{equation}
Here the monomer index $n$ is a continuous variable ranging from 0 to the total number of bonds $N$.  

Our analysis focuses on the static correlation functions of the Rouse modes $p$ and $q$.  In the continuum limit these functions are given by
\begin{eqnarray}
\lefteqn{\rousecorr_{pq} = \langle \vec{X}_p \cdot \vec{X}_q \rangle} \nonumber \\
& = & \frac{1}{N^2} \int_{0}^N  \D n \int_{0}^N \D m \, 
\big \langle \vec{r}(n) \cdot \vec{r}(m) \big \rangle \cos\frac {np\pi}{N} \cos\frac {mq\pi}{N} 
\nonumber \\
& = & - \frac{1}{2N^2} \int_{0}^N \!\! \D n \int_{0}^N \!\!\D m \, 
\big \langle [\vec{r}(n) - \vec{r}(m) ]^2 \big \rangle \times \nonumber \\
& & \hspace*{45mm} \cos\frac {np\pi}{N} \cos\frac {mq\pi}{N} \;.
\label{eq:rousecorr2}
\end{eqnarray}
In dense melts it is commonly assumed that intrachain and interchain excluded volume interactions compensate each other down to the scale of a monomer \cite{DoiEdwards} so that a mean-field picture should apply.  Static equilibrium features can be obtained by treating the polymer liquid as an ensemble of independent chains displaying ideal random-walk-like conformations.  If this was true, single-chain models, i.e.\ phantom chains, should suffice to fully describe chain conformations.  For (some) phantom chain models, the correlation function of the discrete Rouse modes can be calculated.  Therefore, it should be possible to predict, say, the $p$-dependence of the diagonal elements of the simulated Rouse mode matrix, $\rousecorr_{pp}$, from an appropriate phantom chain model.  In the following we want to provide evidence against this expectation.

The simplest phantom chain model is a freely-jointed chain (FJC) model.  For the FJC model the result for $\rousecorr_{pq}$, obtained with the discrete Rouse modes [\eref{eq:rousemodedis}], reads (see e.g.\ \cite{Verdier1966})
\begin{multline}
\rousecorr_{pq} = \delta_{pq}\, \frac{\be^2}{8N}\left[\frac{1}{
\sin (p\pi/2N)}\right]^{2} \\
\xrightarrow{p/N \ll 1} \;
\delta_{pq}\, \frac{1}{2\pi^2} \frac{N\be^2}{p^2} \qquad (\mbox{for $p>0$}) \;.
\label{eq:rousecorrRW}
\end{multline}
The limit $p/N \ll 1$ coincides with the result obtained for the continuum model \cite{DoiEdwards}.  In \eref{eq:rousecorrRW} $\be$ denotes the effective bond length defined by $\be^2 = \Ree^2/N$ from the end-to-end distance $\Ree$ of asymptotically long chains \cite{DoiEdwards}.

\begin{figure}
\begin{center}
\includegraphics*[width=1.0\linewidth]{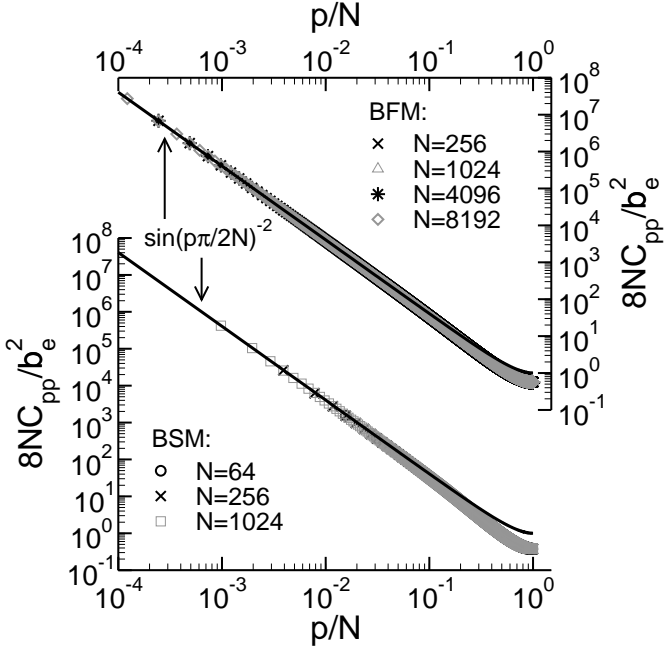}
\end{center}
\caption{Correlation function of the Rouse modes $\rousecorr_{pp}$ versus $p/N$ for the BSM (left ordinate) and the BFM (right ordinate).  All simulation results shown refer to chains with full excluded volume between the monomers.  Following \eref{eq:rousecorrRW} the ordinate is scaled by $8N/\be^2$ (BSM: $\be=1.338$, BFM: $\be=3.244$ \cite{WittmerEtal:PRE2007}).  The solid lines indicate the predictions of the FJC model [\eref{eq:rousecorrRW}].} 
\label{fig:BSM+BFM_rawdata}
\end{figure}

\Fref{fig:BSM+BFM_rawdata} compares \eref{eq:rousecorrRW} to simulation data for excluded volume chains of the BSM and the BFM.  The data for all chain lengths collapse onto a common curve which appears to agree well with $[\sin(p\pi/2N) ]^{-2}$, if $p/N \lesssim 0.01$. (In fact, the agreement is not as good as it seems; we return to this point below.) For larger $p/N$, however, deviations occur. The FJC model overestimates the correlation, especially for $p/N \gtrsim 0.1$.  In \cite{KreerBaschnagel2001} and more recently also in \cite{MolinEtal:JPCM2006} it was argued that these modes are dominated by the microstructure of the simulated chain model.  For instance, $p/N \gtrsim 0.3$ corresponds to subunits of a trimer and smaller. On these local scales, the fact that the repulsive monomer interactions of the BSM and the BFM avoid immediate backfolding of the chain and thus confer some intrinsic stiffness to the polymer should be taken into account.

The simplest way to achieve this consists in replacing the FJC by a (generalized) freely-rotating chain (FRC).  One can introduce the bond angle as a further degree of freedom and still carry out the summation to determine $\rousecorr_{pp}$.  For $N \gg 1$ the result reads \cite{KreerBaschnagel2001}
\begin{multline}
\frac{8N}{\be^2} \rousecorr_{pp} = \left[\frac{1}{\sin (p\pi/2N)}\right]^{2} + \\
\frac{4 \alpha}{1+2\alpha\cos(p\pi/N)+ \alpha^2} \; ,
\label{eq:rousecorrGFRC}
\end{multline}
where $\alpha = \langle \cos \theta \rangle$.
  
\begin{figure}
\begin{center}
\includegraphics*[width=0.9\linewidth]{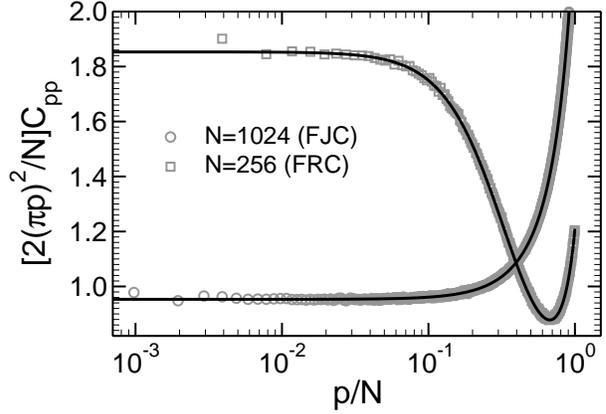}
\end{center}
\caption{Correlation function of the Rouse modes $\rousecorr_{pp}$ versus $p/N$ for BSM phantom chains (FJCs and FRCs, cf.\ Section~\ref{subsubsec:hard2soft}) of lengths $N=256$ and $N=1024$.  $\rousecorr_{pp}$ is scaled by $2(\pi p)^2/N$ so that the ordinate tends to $\be^2$ for $p/N \rightarrow 0$.  The simulation results for FJCs and FRCs are compared to Eqs.~\eqref{eq:rousecorrRW} and \eqref{eq:rousecorrGFRC}, respectively.  These comparisons utilize the following parameters:  for FJCs, $\be^2 = 0.9532$ and for FRCs, $\be^2 = C_\infty \times 0.9532$, where the characteristic ratio is given by $C_\infty =(1-\alpha)/(1+ \alpha)$ (with $\alpha = -0.320713$).  The values of $\be$ and $\alpha$ are determined in the simulation. } 
\label{fig:phantomchains}
\end{figure}

\begin{figure}
\begin{center}
\includegraphics*[width=0.9\linewidth]{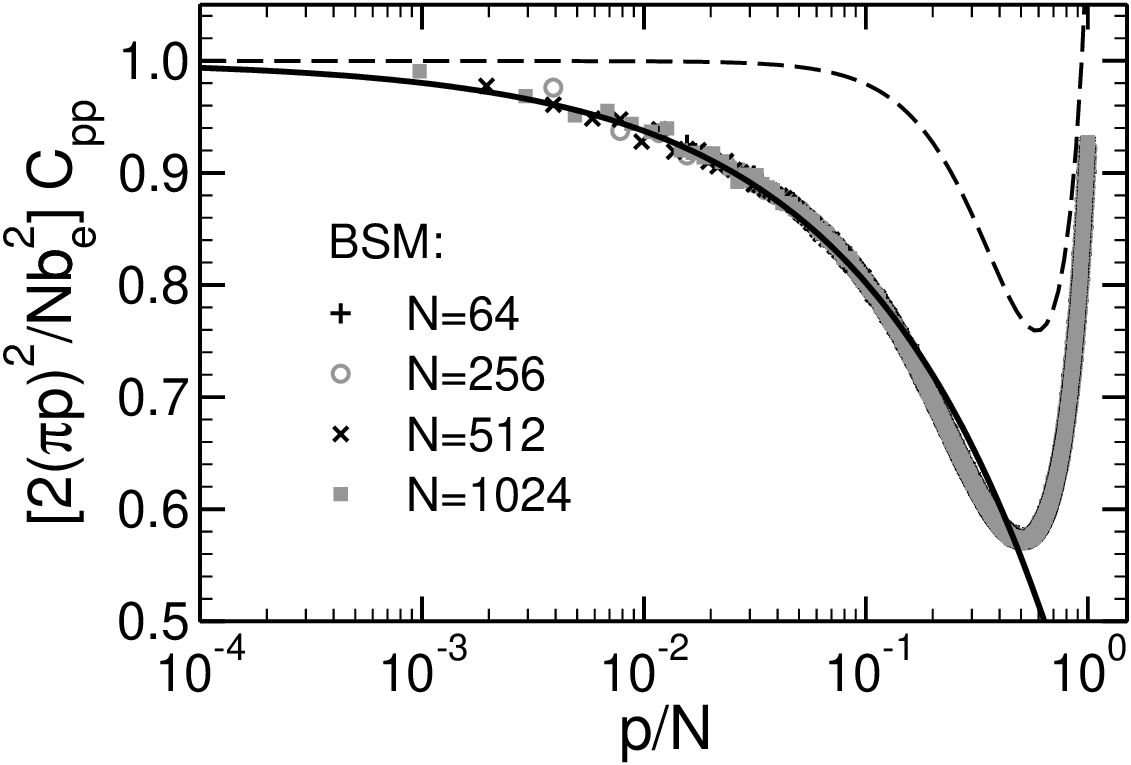}
\includegraphics*[width=0.9\linewidth]{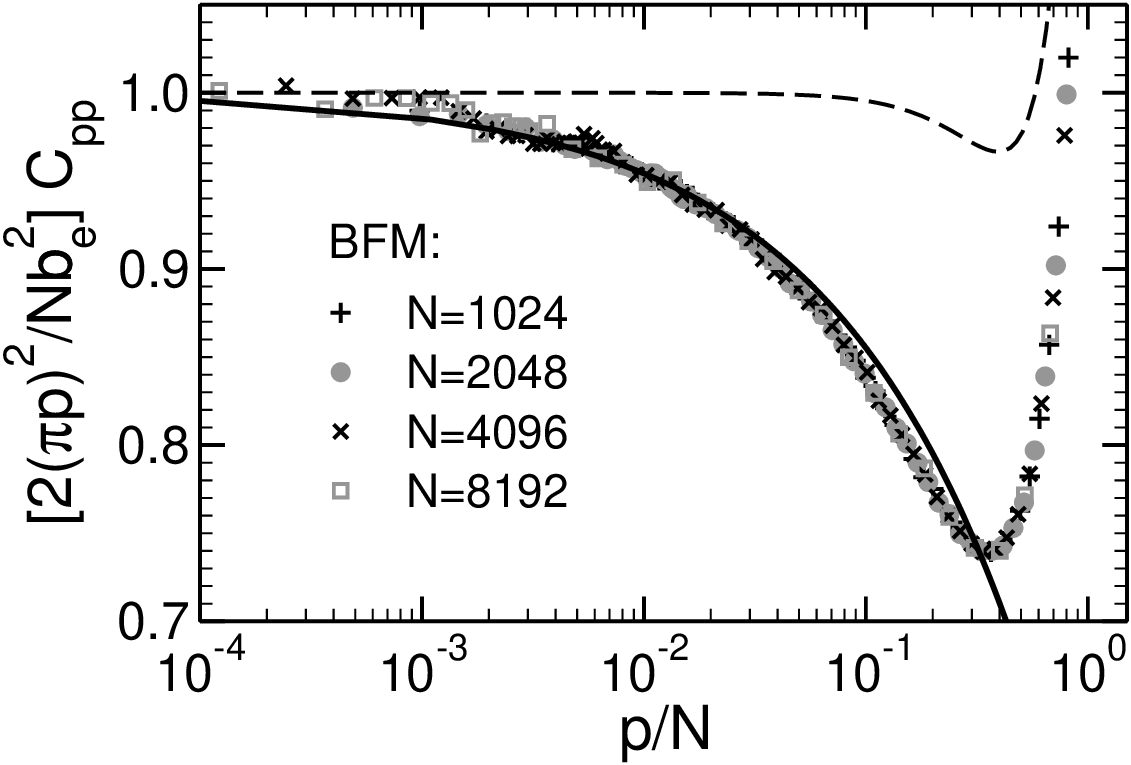}
\end{center}
\caption{Upper panel: Correlation function of the Rouse modes $\rousecorr_{pp}$ versus $p/N$ for BSM excluded volume chains.  $\rousecorr_{pp}$ is scaled by $2(\pi p)^2/N\be^2$ so that the ordinate tends to $1$ for $p/N \rightarrow 0$ ($\be=1.338$ \cite{WittmerEtal:PRE2007}).  The solid line presents \eref{eq:rousecorr_long_range} with parameters from Table~\ref{tab:BSM}.  The dashed line indicates the prediction of the FRC model, \eref{eq:rousecorrGFRC}, with $\alpha=-0.1948$ (measured in the simulation).  Lower panel: Same as in the upper panel, but for BFM excluded volume chains with $\be=3.244$ (Table~\ref{tab:BSM}) and $\alpha=- 0.1055$ (measured in the simulation).}
\label{fig:BSMchains}
\end{figure}

\Fref{fig:phantomchains} shows that Eqs.~\eqref{eq:rousecorrRW} and \eqref{eq:rousecorrGFRC} agree respectively with the simulation results for the corresponding phantom chain models.  However, they do not for the full-excluded-volume chains in the melt.  This is demonstrated in \fref{fig:BSM+BFM_rawdata} for the FJC model and in \fref{fig:BSMchains} for the FRC model.  To emphasize the differences between the numerical results and FRC prediction the simulation data are divided by the asymptotic behavior expected (and found) for $p/N \ll 1$ [\eref{eq:rousecorrRW}].  With increasing $p/N$, the simulation data show a continuous depression, which is not reproduced by the FRC model.  (The FJC  model has not even a minimum; cf. \fref{fig:phantomchains}.)  This depression is, however, compatible with a theory that accounts for the impact of residual excluded volume interactions in the melt.  We will sketch this theory in the next section.   

\section{Static Rouse mode correlations: corrections to chain ideality}
\label{sec:fullexvol}
By means of the random-phase approximation (RPA) Edwards derived the pair potential between two monomers in a dense, three-dimensional multi-chain system of asymptotically long chains \cite{DoiEdwards,Edwards_JPhysA1975},
\begin{equation}
\widetilde{v}(r) = v \left (\delta(r) - \frac{\exp(-r/\xi)}{4\pi r \xi^2} \right ) \;.
\label{eq:edwardspot}
\end{equation}
This potential consists of two terms.  The first term represents the bare repulsive interaction between two monomers.  It is of very short range and characterized by the excluded volume parameter $v$.  The latter is related to the compressibility of the multi-chain system [see \eref{eq:fcap3}].  The second term in \eref{eq:edwardspot} results from the repulsion of all monomers in the system whose compound effect is to attenuate the bare interaction between the tagged monomer pair.  It is of range $\xi$, where $\xi = b/ (12 \rho v)^{1/2}$ is the (Edwards) correlation length of collective density fluctuations, and $b$ denotes the effective bond length of an ideal polymer chain with all interactions switched off ($v=0$).
 
In a dense (three-dimensional \cite{SemenovJohner2003}) system we expect both terms to nearly compensate each other and $\widetilde{v}(r)$ to be small.  This suggests that a first-order perturbation calculation should be appropriate to explore the influence of $\widetilde{v}(r)$ on the conformational properties of a polymer melt.  Following \eref{eq:rousecorr2} $\rousecorr_{pq}$ can be obtained from the mean-square end-to-end distance between monomers $n$ and $m$, $\langle [\vec{r}_n - \vec{r}_m ]^2 \rangle$.  For $\langle [\vec{r}_n - \vec{r}_m ]^2 \rangle$ the perturbation calculation has been carried out.  The result for infinitely long chains reads ($s=n-m$) \cite{WittmerEtal:PRE2007}
\begin{equation}
\Ree^2(s) = \be^2 \Big [|s| - \ke \sqrt{|s|}\Big ]\,, \quad 
\ke = \sqrt{\frac{24}{\pi^3}} \, \frac{1}{\rho \be^3}\;.
\label{eq:lr1}
\end{equation}
We see that $\widetilde{v}(r)$ corrects the result for ideal chains, $\Ree^2(s) = b^2 s$, in two ways.  It increases the effective bond length from $b$ to $\be$, as already predicted by Edwards \cite{DoiEdwards}, and leads to a swelling of the internal distance between two monomers due to the second term $ - \ke\sqrt{s}$.  (We refer to $\ke$ as `swelling factor' in the following.)  This swelling is responsable for the deviations of $\rousecorr_{pp}$ from the ideal behavior (cf.\ \fref{fig:BSMchains}), as we will demonstrate now.
  
$\Ree^2(s)$ depends only on the absolute value of the curvilinear distance $s$.  Thus, we first rewrite \eref{eq:rousecorr2} as 
\begin{eqnarray}
\lefteqn{\rousecorr_{pq}} \nonumber \\
& = & - \frac{1}{2N^2} \int_{0}^N \D n \int_{0}^N \D m \, 
\Ree^2(|n - m|) \cos\frac {np\pi}{N} \cos\frac {mq\pi}{N} \nonumber \\
& = & - \frac{1 + (-1)^{p+q}}{2N^2} \int_{0}^N \D n \int_{0}^n \D m \, \Ree^2(n - m) \times \nonumber \\
& & \hspace*{45mm} \cos\frac {np\pi}{N} \cos\frac {mq\pi}{N} \;,
\label{eq:lr2}
\end{eqnarray}
and then insert \eref{eq:lr1} into \eref{eq:lr2}.  The first term of \eref{eq:lr1} will yield the result for ideal chains, i.e.\ \eref{eq:rousecorrRW} in the limit $p/N \ll 1$.  The second term of \eref{eq:lr1} will modify \eref{eq:rousecorrRW} in two respects: (i) it provides a correction to ideal chain behavior (i.e.\ to the diagonal terms of $\rousecorr_{pq}$), (ii) the Rouse modes are not diagonal any longer.

To substantiate these expectations we have to calculate the integral:
\begin{equation}
I_{pq} = \int_{0}^N \D n \int_{0}^n \D m \, \sqrt{n-m} \cos\frac {np\pi}{N} 
\cos\frac {mq\pi}{N} \;.
\label{eq:lr3}
\end{equation}
Substituting $s=n-m$ and integration by parts gives
\begin{multline}
I_{pq} = - \frac{1}{\sqrt{2}\pi} \bigg(\frac{N}{q}\bigg)^{3/2} \int_{0}^N \D s
\cos\frac {sp\pi}{N} \\ 
\bigg \{\cos\frac {sq\pi}{N} S\bigg(\sqrt{\frac {sq\pi}{N}}\bigg)
- \sin\frac {sq\pi}{N} C\bigg(\sqrt{\frac {sq\pi}{N}}\bigg) \bigg \} \;,
\label{eq:lr4}
\end{multline}
where $S(\sqrt{x})$ and $C(\sqrt{x})$ are the Fresnel integrals
\begin{equation}
\begin{aligned}
S(\sqrt{x}) 
& = \frac{1}{\sqrt{2\pi}} \int_0^x \D y \, \frac{\sin y}{\sqrt{y}} \;, \\
C(\sqrt{x}) 
& = \frac{1}{\sqrt{2\pi}} \int_0^x \D y \, \frac{\cos y}{\sqrt{y}} \;.
\end{aligned}
\label{eq:lr5}
\end{equation}
For the term in curly braces in \eref{eq:lr4} we may use the expression
\begin{multline}
\cos\frac {sq\pi}{N} S\bigg(\sqrt{\frac {sq\pi}{N}}\bigg)
- \sin\frac {sq\pi}{N} C\bigg(\sqrt{\frac {sq\pi}{N}}\bigg) =\\
 \frac 12 \bigg [\cos\frac {sq\pi}{N} - \underline{\sin\frac {sq\pi}{N}} \bigg ] 
- \sqrt{\frac{2}{\pi}} \int_0^\infty \D t \, \E^{-2 \sqrt{\frac{sq\pi}{N}}\,t} \cos t^2 \;.
\label{eq:lr6}
\end{multline} 
This expression is helpful because the underlined term vanishes upon integration over $s$, whereas the other two terms provide the expected corrections to ideal behavior: the cosine term amends the diagonal components and the third term of \eref{eq:lr6} makes the Rouse modes nondiagonal.

Putting these results together we find (for $p,q > 0$)
\begin{align}
\rousecorr_{pq} 
& = \frac{1}{2\pi^2} \frac{\be^2}{N} \bigg (\frac{N}{p}\bigg)^2 \delta_{pq} 
\quad (\mbox{ideal}) \nonumber \\
& - \frac{1}{2\pi^2} \frac{\be^2}{N} \, \frac{\pi \ke}{\sqrt{8}}
\bigg (\frac{N}{q}\bigg)^{\frac{3}{2}} \bigg \{\delta_{pq} - 
\underline{\frac{[1 + (-1)^{p+q}] \sqrt{8}}{\pi^{3/2}p} \, \times} \nonumber \\
& \hspace*{5mm} \underline{\int_0^{p\pi} \D x \cos x \int_0^\infty \D t \, 
\E^{-2 \sqrt{\frac{q}{p}\,x} \, t} \cos t^2} \bigg \} \;.
\label{eq:lr7}
\end{align}
If we now approximate the integral of the underlined term by replacing the upper bound $p\pi$ by $\infty$, we obtain
\begin{multline}
\int_0^{\infty} \D x \cos x \int_0^\infty \D t \, 
\E^{-2 \sqrt{\frac{q}{p}\,x} \, t} \cos t^2 \\
\begin{aligned}
& = \sqrt{\frac{\pi}{8}} \, \frac{q/p}{[(q/p) + 1][\sqrt{q/p} + 1]} \\
& = \sqrt{\frac{\pi}{8}} \, q^{3/2}p \; \frac{p^{-1/2} - q^{-1/2} }{q^2-p^2} \;,
\end{aligned}
\label{eq:lr8}
\end{multline}
and the underlined term vanishes in the limit $p \rightarrow \infty$.  

This approximation suggests the Rouse modes to remain essentially diagonal despite of excluded volume interactions.  Indeed, numerical analysis of the theoretical result and inspection of the simulation data reveal that the nondiagonal terms $\rousecorr_{pq}$ are smaller than the self-correlation $\rousecorr_{pp}$ by at least two orders of magnitude.  The small values of $\rousecorr_{pq}$ make it difficult to separate signal from noise in the simulation.  However, the results are still indicative of the parity property predicted by \eref{eq:lr7}: modes with `$p+q=$ odd' vanish, while those with `$p+q=$ even' are finite.  This parity is a consequence of the translational invariance---$\Ree$ depends only on $|s|$---of the chain which we assumed to be infinitely long in \eref{eq:lr1}.  A more quantitative analysis of the cross correlations requires improvement of the numerical precision and perhaps refinement of the theory.  Two refinements are envisageable:  It is possible to relax the approximation `$p=\infty$' in the underlined integral of \eref{eq:lr7} and along with that, to also account for finite-$N$ effects (the latter has been done before in the discussion of the form factor \cite{BeckrichEtal:Macro2007}).  We plan to work on both aspects---improvement of numerical precision in the simulation and refinement of the theory---in the future.  


Here we continue with the approximation `$p=\infty$' and examine its consequences. For the diagonal Rouse modes we find 
\begin{equation}
\rousecorr_{pp} = \frac{N\be^2}{2(\pi p)^2} \bigg [ 1 -
\frac{\pi}{\sqrt{8}} \, \ke \, \sqrt{\frac{p}{N}}\, \bigg ]
\qquad (\mbox{for $p>0$}) \; .
\label{eq:rousecorr_long_range}
\end{equation}


\begin{table}
\caption{\label{tab:BSM}Survey of BSM and BFM parameters for systems with variable monomer overlap.  $E$ is the energy penalty for full monomer overlap.  $g$ and $\be(\sblob)$ denote respectively the number of monomers in the blob and the statistical segment length.  $\kone$ is the empirical swelling factor discussed in \cite{WittmerEtal:PRE2007}.  For the BSM we take $\kone = 1.3 \ke$ and for the BFM, $\kone = \ke$ [$\ke$ is defined in \eref{eq:lr1}].}
\begin{center}
{\scriptsize  
\begin{tabular}{cccc|cccc}
\hline
\hline\\[-2mm]
\multicolumn{4}{c|}{BSM} & \multicolumn{4}{|c}{BFM} \\
\hline\\[-2mm]
$E$ & $\sblob$ & $\be(\sblob)$ & $\kone$ & $E$ & $\sblob$ & $\be(\sblob)$ & $\kone$ \\[1mm]
\hline \hline\\[-2mm]
$\infty$ & 0.08 & 1.338 & 0.564 & $\infty$ & 0.246 & 3.244 & 0.412 \\
4.68     & 0.23 & 1.306 & 0.607 & 3      &    0.85 & 3.213 & 0.424 \\ 
1.94     & 0.51 & 1.274 & 0.654 & 0.5    &    4.43 & 3.055 & 0.494 \\
0.5      & 1.83 & 1.208 & 0.767 & 0.1    &   20    & 2.917 & 0.567 \\
0.1      & 9    & 1.129 & 0.939 & 0.01   &  200    & 2.795 & 0.645 \\
0.02     & 45   & 1.073 & 1.094 & 0.001  & 2000    & 2.740 & 0.679 \\[1mm]
\hline
\hline
\end{tabular}
}
\end{center}
\end{table}

This result is compared to the simulation data of the BSM and the BFM in \fref{fig:BSMchains}.  For this comparison we used the empirical swelling factors $\kone$ determined earlier in an analysis of the segmental size distribution in polymer melts \cite{WittmerEtal:PRE2007}.  This analysis revealed that $\kone$ agrees with the theoretical value $\ke$ for the BFM, while for the BFM, a slightly larger value had to be used, $\kone = 1.3 \ke$ (Table~\ref{tab:BSM}).  Here we adopt these values so that the comparison between theory and simulation contains no further adjustable parameter.  \Fref{fig:BSMchains} demonstrates that the theory can indeed account for the systematic depression of $\rousecorr_{pp}$ below the ideal asymptote, if $p/N \lesssim 0.1$.  On the other hand, the behavior of $\rousecorr_{pp}$ for larger values of $p/N$ and in particular the upturn for $p \rightarrow N$ cannot be described because such large modes should be strongly influenced by the microstructure of the polymer model, which is not treated correctly by the present theory.

\section{Chains with variable monomer overlap}
\label{sec:finiteoverlap}

\begin{figure}
\begin{center}
\includegraphics*[width=1.\linewidth]{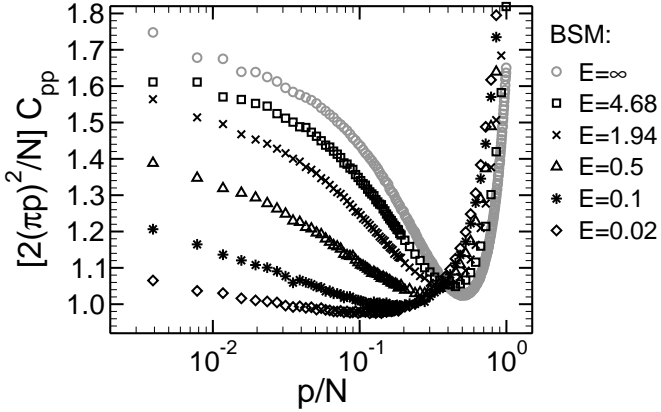}
\end{center}
\caption{$\rousecorr_{pp}$ versus $p/N$ for hard ($E=\infty$) and soft (finite values for $E$) monomers from BSM simulations for $N=256$. The smaller the value of the overlap energy $E$, the softer the monomers (see Sec.~\ref{sec:models}).}
\label{fig:diag_rouse_forcecap_rawdata}
\end{figure}

In section~\ref{sec:rouse} we found that phantom chain models are not capable of describing conformational properties of flexible polymers in a melt by comparing simulation data for excluded volume chains to phantom chain calculations.  The simulation provides a further means to support this conclusion.  One can vary the strength of the repulsive interaction between two monomers. 

\Fref{fig:diag_rouse_forcecap_rawdata} exemplifies the results of such an analysis for the BSM.  The figure depicts $\rousecorr_{pp}$ for chains with full excluded volume ($E=\infty$) and with finite energy penalty $E$ for monomer overlap.  Obviously, the $p$-dependence of $\rousecorr_{pp}$ qualitatively changes with $E$.  As the monomers become softer with decreasing $E$, the curvature of $\rousecorr_{pp}$ for $p/N \lesssim 0.1$, observed for the excluded volume chains, progressively vanishes.  This clearly demonstrates that this feature is determined by the strength of monomer repulsion in the melt.  It is possible to rationalize these findings by an extension of the perturbation theory of Section~\ref{sec:fullexvol} to soft monomers. 

\subsection{Internal distances for finite monomer overlap}
 To calculate $\rousecorr_{pp}$ we begin, as before, by determining $\Ree^2(s)$ first. Quite generally, $\Ree^2(s)$ is related to the bond correlation function
\begin{equation}
P_1(|n-m|) = \bigg \langle \frac{\partial \vec{r}(n)}{\partial n} \cdot 
\frac{\partial \vec{r}(m)}{\partial m} \bigg \rangle \;
\label{eq:fcap0}
\end{equation}
by
\begin{align}
\Ree^2(s) 
& = \int_0^s \D n \int_0^s \D m \, \bigg \langle 
\frac{\partial \vec{r}(n)}{\partial n} \cdot \frac{\partial \vec{r}(m)}{\partial m} 
\bigg \rangle \nonumber \\ 
& = 2 \blength^2 s \int_0^s \D t \,\Big [ 1 - \frac{t}{s} \Big ] P_1(t) \;,
\label{eq:fcap1}
\end{align}
where $\blength$ is the average bond length.  For monomers with different degree of softness the expression for $P_1(s)$ has recently been obtained from first-order perturbation theory.  It reads \cite{WittmerEtal:preprint2007}
\begin{equation}
P_1(x) = \frac{4 \sqrt{\pi} \kp(\sblob)}{\sblob^{3/2}} \bigg [ \frac{1}{\sqrt{\pi x}} 
- \sqrt{2} \E^{2x} \operatorname{erfc}\big( \sqrt{2x} \big) \bigg ] \;,
\label{eq:fcap4}
\end{equation}
where 
\begin{equation}
\kp(\sblob) = \frac{1}{8} \, \ke(\sblob) \, 
\bigg (\frac{\be(\sblob)}{\blength(\sblob)} \bigg)^2 \;,
\label{eq:fcap2a}
\end{equation}
$\ke(\sblob)$ is given by \eref{eq:lr1} with $\be=\be(\sblob)$, `$\operatorname{erfc}$' is the complementary error function, $x=s/\sblob$, and $\sblob$ is the number of monomers in a blob.  (We will comment on the blob after \eref{eq:fcap6}.)  $\sblob$ is related to the correlation length $\xi$ and the effective bond length $b$ of an unperturbed ideal chain via \cite{WittmerEtal:preprint2007}
\begin{equation}
\sblob = 12 \bigg (\frac{\xi}{b} \bigg )^2 = \frac{1}{v\rho} = S(q\rightarrow 0) \; ,
\label{eq:fcap3}
\end{equation}
where $S(q)$ is the collective structure factor of the polymer melt.  Inserting \eref{eq:fcap4} into \eref{eq:fcap1} we obtain
\begin{eqnarray}
\lefteqn{1 - \frac{\Ree^2(s,g)}{\be^2(g) s} =} \nonumber \\
& & \frac{\ke(\sblob)}{\sqrt{\sblob}}
\left [ \frac{1}{\sqrt{x}} 
- \sqrt{\frac{\pi}{8}}\, \frac{1}{x} \Big \{ 1- \E^{2x} 
\operatorname{erfc}\big( \sqrt{2x} \big) \Big \}
\right ] \; .
\label{eq:fcap5}
\end{eqnarray}
\Eref{eq:fcap5} has the following asymptotic behavior
\begin{multline}
\frac{\sqrt{\sblob}}{\ke(\sblob)}\bigg [1- \frac{\Ree^2(s,\sblob)}{\be^2(\sblob)s} 
\bigg ]\\
\simeq  
\begin{cases}
\sqrt{\frac{\displaystyle \pi}{\displaystyle 2}} \left (1 - \frac{\displaystyle 2^{5/2}}{\displaystyle 3\pi^{1/2}} \sqrt{x} \right) & \text{for $x \ll 1$} 
\;, \\[2mm]
\frac{\displaystyle 1}{\displaystyle \sqrt{x}} & \text{for $x \gg 1$} \;.
\end{cases}
\label{eq:fcap6}
\end{multline}

Here we return to the comment mentioned above.  Softening of the monomer repulsion introduces the blob size $\sblob$ as a new parameter, and the situation becomes similar to that of semidilute solutions in good solvent.  Inside the blob chain segments behave as if they were in dilute solution.  They do not see their neighbors and are slightly swollen.  Since the monomer-monomer repulsion is weak, we find for $x \ll 1$ that the swelling takes the form of a first-order Fixman expansion, familiar from the study of excluded volume effects in dilute solutions close to the theta point \cite{DoiEdwards,RubinsteinColby}.  For large $x$---that is, for chain segments much bigger than the blob size---\eref{eq:fcap5} gives back \eref{eq:lr1}.  On such large scales the polymer system behaves like a dense melt of blob chains which repel each other.  Hence, the swelling of internal distances is the same as that of chains whose monomers have full excluded interaction.  

\begin{figure}
\begin{center}
\includegraphics*[width=0.9\linewidth]{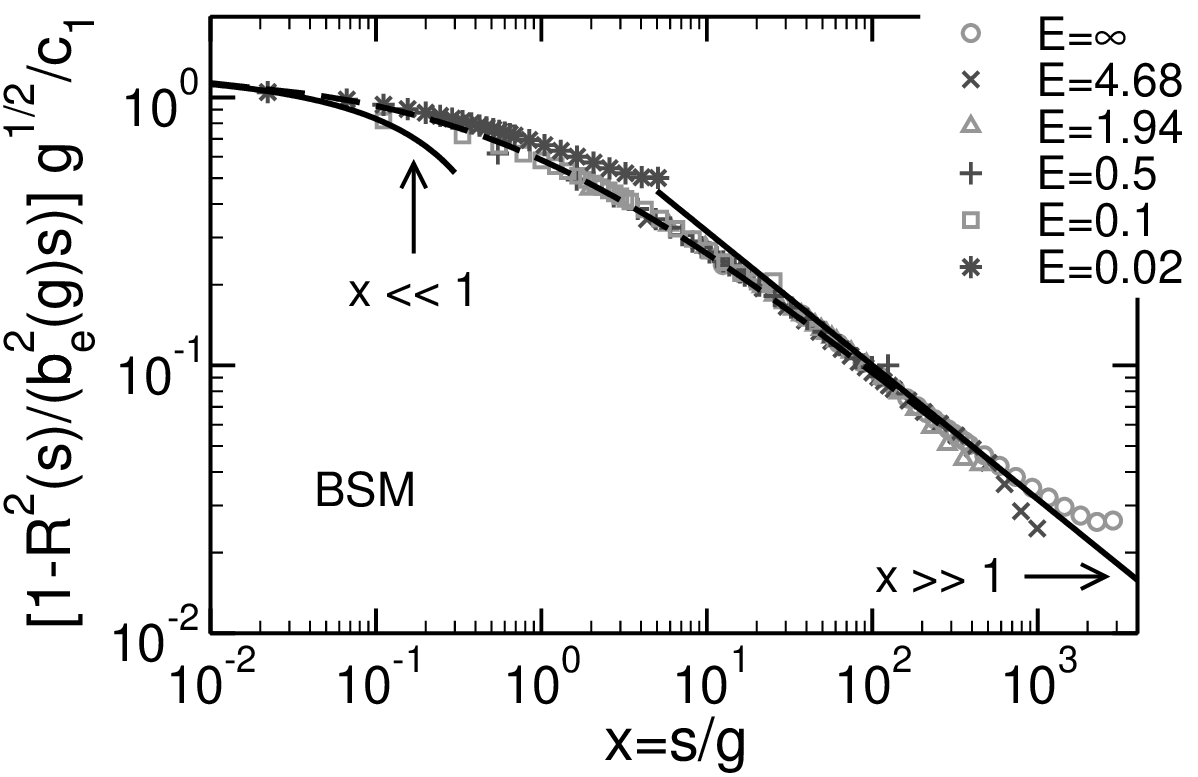}
\includegraphics*[width=0.9\linewidth]{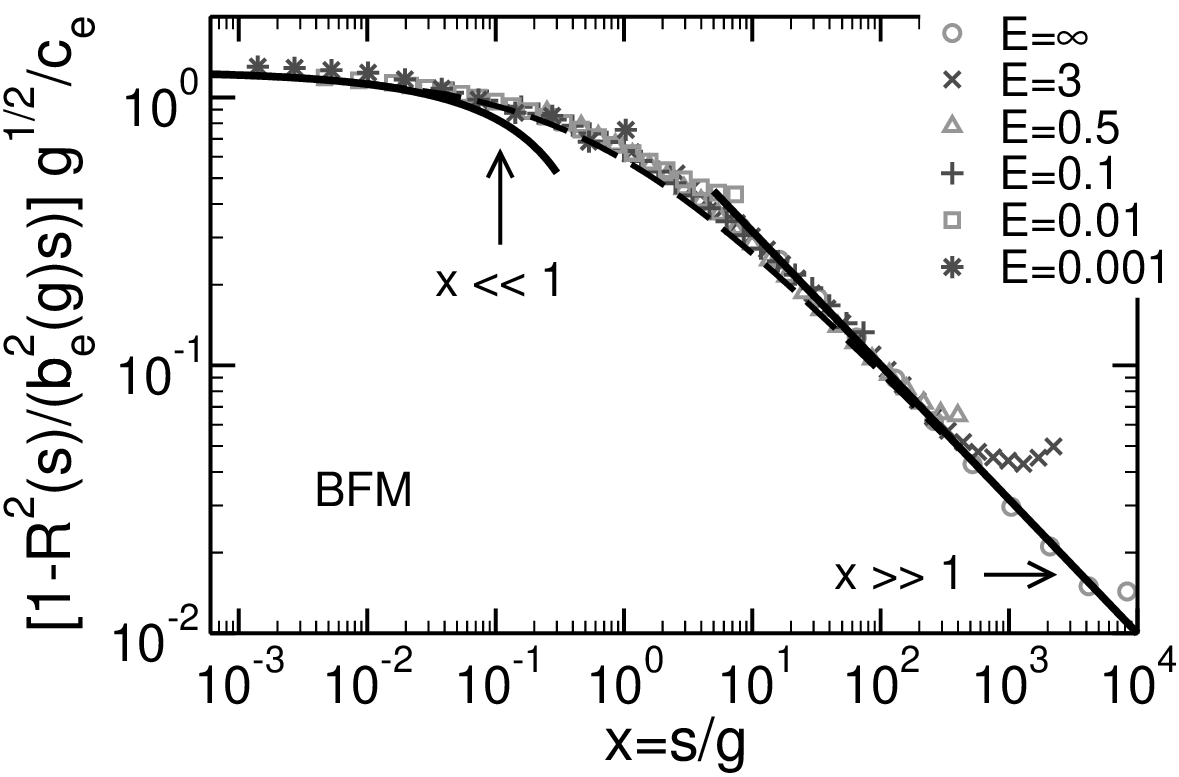}
\end{center}
\caption{Scaling plot of $\Ree^2(s)$, as suggested by \eref{eq:fcap5}, for monomers of different softness.  The upper panel depicts BSM data for $N=256$, the lower panel shows BFM data for $N=2048$.  The dashed line indicates \eref{eq:fcap5}.  The solid lines show the asymptotic behavior from \eref{eq:fcap6} for small and large $x=s/g$.  The values used to scale the axes may be found in Table~\ref{tab:BSM}.}
\label{fig:odfintra_and_r2s_N256_forcecap}
\end{figure}

To compare simulation and theory we have to determine the swelling factor and the blob size.  The blob size was obtained from the low-$q$ limit of $S(q)$ [cf.\ \eref{eq:fcap3}].  For large $E$ this limit can be read off reliably from the simulation data.  For the weakest energy penalties, however, $S(q)$ does not reach a plateau for the smallest $q$-values studied.  Here we determined $g$ by fitting $S(q)$ to the RPA formula $S(q)^{-1} = g^{-1} + \be^2(g) q^2/12$ \cite{DoiEdwards}.  The resulting blob sizes have thus larger error bars than those for large $E$.  The fit to the RPA formula also yields an estimate for $\be(g)$.  This estimate can be crosschecked and optimized when determining the swelling factors.  For the swelling factors we adopt the result found previously---that is, $\kone=\ke$ for the BFM and $\kone=1.3\ke$ for the BSM---but allow $\kone$ to depend on $\sblob$ via $\be$.  For small $\sblob$, i.e.\ weak to vanishing monomer overlap, $\be^2(\sblob)$ may be obtained fairly reliably by fitting the asymptotic behavior for $x \gg 1$ to the simulation data for $\Ree^2(s,\sblob)$ \cite{WittmerEtal:PRE2007}.  For large $\sblob$, this fitting procedure is more problematic because one has to choose the fit interval, and it is hard to find an extended regime (of intermediate $x$ values) where the simulation and theoretical curves have the same shape.  The results for $\be^2(\sblob)$ were thus obtained by a two-step procedure: First, \eref{eq:fcap5} was fitted to $\Ree^2(s,\sblob)$.  Then, the fit result for $\be^2(\sblob)$ was optimized in such a way that it yields a good data collapse in \fref{fig:odfintra_and_r2s_N256_forcecap} and that the resulting master curve is close to the theoretical prediction, \eref{eq:fcap5}.  The so-obtained values for $g$ and $\be^2(g)$ are collected in Table~\ref{tab:BSM}.

In \fref{fig:odfintra_and_r2s_N256_forcecap} we compare Eqs.~\eqref{eq:fcap5} and \eqref{eq:fcap6} with the numerical results from the BSM and the BFM for various values of the energy penalty $E$.  The figure demonstrates that there is good agreement between theory and simulation, perhaps with the exception of the smallest $E$-values.  However, here the numerical uncertainties for $g$ and $\be$ are largest.  

\subsection{Rouse modes for finite monomer overlap}
\begin{figure}
\begin{center}
\includegraphics*[width=0.9\linewidth]{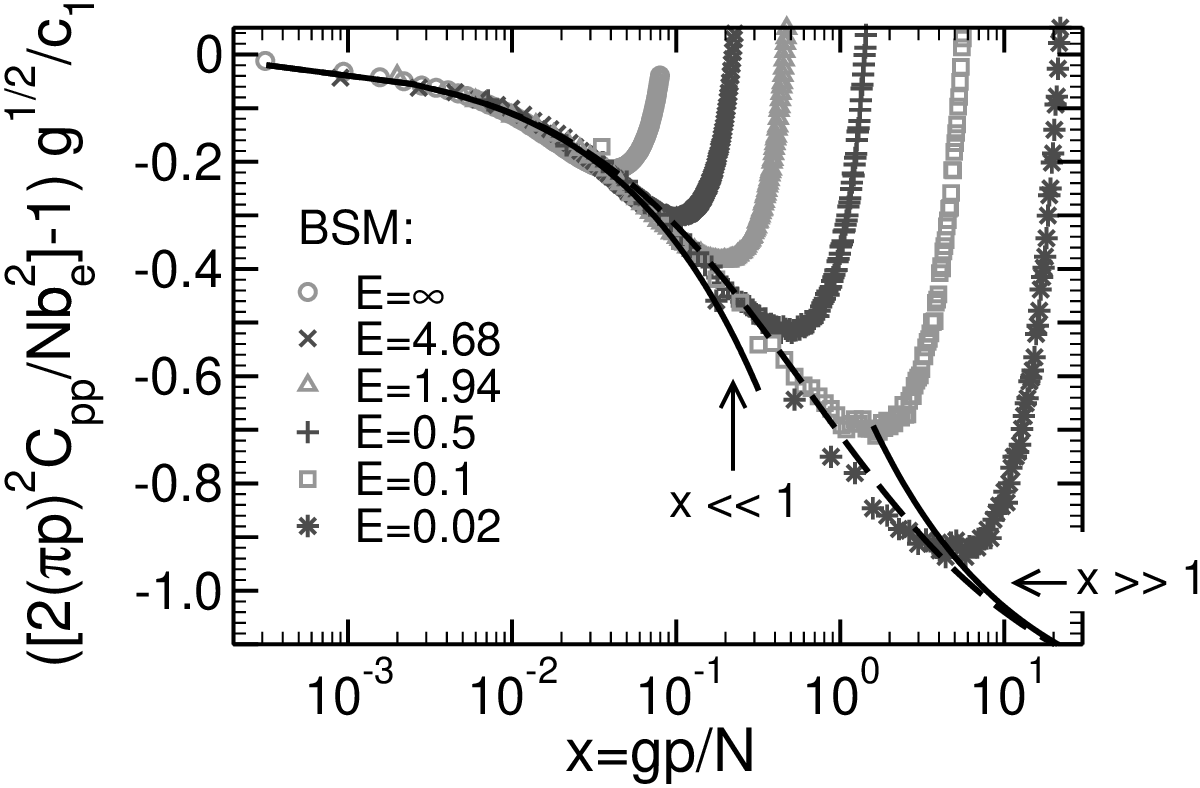}
\includegraphics*[width=0.9\linewidth]{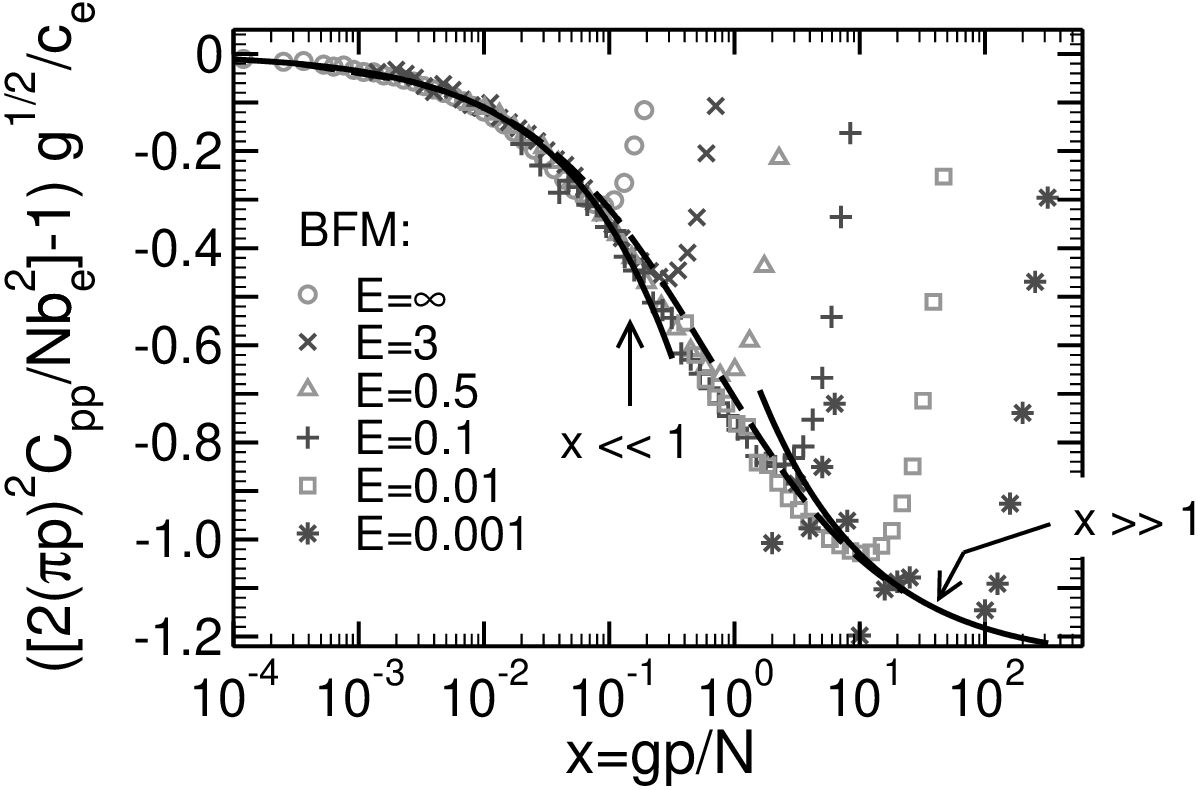}
\end{center}
\caption{Scaling plot of $\rousecorr_{pp}$, as suggested by \eref{eq:fcap7}, for monomers of different softness.  The upper panel depicts BSM data for $N=256$, the lower panel shows BFM data for $N=2048$.  The dashed line indicates \eref{eq:fcap7}.  The solid lines show the asymptotic behavior from \eref{eq:fcap8} for small and large $x=gp/N$.  The values to scale the axes are given in Table~\ref{tab:BSM}.
}
\label{fig:diag_rouse_N256_forcecap}
\end{figure}

The result for the mean-square internal end-to-end distance may be inserted into \eref{eq:rousecorr2} to derive a scaling prediction for the correlation function of the Rouse modes.  We find for the diagonal elements of the Rouse mode matrix the following expression (again in the limit $p \rightarrow \infty$, see approximation in \eref{eq:lr8} for comparison) 
\begin{multline}
\frac{2(\pi p)^2}{N \be^2(\sblob)} \rousecorr_{pp} - 1 = 
\sqrt{\frac{\pi}{2}}\, \frac{\ke(\sblob)}{\sqrt{\sblob}} \, 
\bigg \{ \frac{1}{1 + 4/(\pi x)^2} \, \times \\
\frac{1}{\sqrt{\pi x}}
\bigg [ 1 - \sqrt{\pi x}  + \frac \pi 2 \,x \bigg ] 
- \frac 12 \, \sqrt{\pi x} \bigg \} \; ,
\label{eq:fcap7}
\end{multline}
where $x=gp/N$.  This equation has the following asymptotic behavior
\begin{multline}
\frac{\displaystyle \sqrt{\sblob}}{\displaystyle \ke(\sblob)}
\bigg [\frac{2(\pi p)^2}{N\be^2(\sblob)} \rousecorr_{pp} - 1 \bigg ]\\
\simeq
\begin{cases}
- \frac{\displaystyle \pi }{\displaystyle \sqrt{8}} \, \sqrt{x}
& \text{for $x \ll 1$} \;, \\[4mm]
\sqrt{\frac{\displaystyle \pi}{\displaystyle 2}} \, \Big [- 1 + 
1/\sqrt{\pi x} \Big ] 
& \text{for $x \gg 1$} \;.
\end{cases}
\label{eq:fcap8}
\end{multline} 
Equations~\eqref{eq:fcap7} and \eqref{eq:fcap8} are compared to BSM and BFM data in \fref{fig:diag_rouse_N256_forcecap}.  The agreement between theory and simulation is of similar quality as in \fref{fig:odfintra_and_r2s_N256_forcecap}.

\section{Summary}
\label{sec:summary}
The simulation models studied in this work have very flexible chains in common.  As chain stiffness effects are reduced to a minimum---they only come in due to the avoidance of immediate chain backfolding---one should expect ideal chain behavior to appear clearly.  

We demonstrated that, even under these favorable circumstances, the assumption of chain ideality on all length scales down to the monomer size does not hold.  The correlation function of the Rouse modes displays systematic deviations from the $(p/N)^{-2}$ scaling expected for ideal chains.  Our analysis suggests that these deviations may be traced back to the fact that repulsive interactions between chain segments in the melt are not fully screened.  For chain segments of size $s \gg 1$ there is an entropic penalty $\sim 1/\sqrt{s}$ for bringing two segments together \cite{WittmerEtal:PRE2007}. This penalty swells the segments and causes systematic deviations from Flory's ideality hypothesis.  The picture of independent chains with random-walk-like conformations is thus not acceptable for polymer melts.  Multi-chain effects reflecting the interplay of chain connectivity and melt incompressibility should be taken into account.

In the present work we discussed the impact of these multi-chain effects on static chain properties.  It is, however, natural to expect that they will also affect the polymer dynamics.  An analysis of this influence is underway.

\acknowledgement{
We are indebted to S. Obukhov for valuable discussions and to the IDRIS (Orsay) for a generous grant of computer time.  Financial support by the IUF, the ESF STIPOMAT programme, and the DFG (grant number KR 2854/1--1) is gratefully acknowledged.}



\begin{thebibliography}{10}

\bibitem{rouse1953}
P.~E. Rouse, J. Chem. Phys. {\bf 21},  1272  (1953).

\bibitem{DoiEdwards}
M. Doi and S.~F. Edwards, {\em The Theory of Polymer Dynamics} (Oxford
  University Press, Oxford, 1986).

\bibitem{RubinsteinColby}
M. Rubinstein and R.~H. Colby, {\em Polymer Physics} (Oxford University Press,
  Oxford, 2003).

\bibitem{McLeish_AdvPhys2002}
T.~C.~B. McLeish, Adv. Phys. {\bf 51},  1379  (2002).

\bibitem{PaulSmith_RPP2004}
W. Paul and G.~D. Smith, Rep. Prog. Phys. {\bf 67},  1117  (2004).

\bibitem{HarnauEtal:EPL1999}
L. Harnau, R.~G. Winkler, and P. Reineker, Europhys. Lett. {\bf 45},  488
  (1999).

\bibitem{KrushevEtal_Macromolecules2002}
S. Krushev, W. Paul, and G.~D. Smith, Macromolecules {\bf 35},  4198  (2002).

\bibitem{BulacuGiessen:JCP2005}
M. Bulacu and E. van~der Giessen, J. Chem. Phys. {\bf 123},  114901  (2005).

\bibitem{KreerBaschnagel2001}
T. Kreer, J. Baschnagel, M. M\"uller, and K. Binder, Macromolecules {\bf 34},
  1105  (2001).

\bibitem{MolinEtal:JPCM2006}
D. Molin, A. Barbieri, and D. Leporini, J. Phys.: Condens. Matter {\bf 18},
  7543  (2006).

\bibitem{RichterMonkenbuschAllgeier1999}
D. Richter {\it et~al.}, J. Chem. Phys. {\bf 111},  6107  (1999).

\bibitem{ArbeEtal:Macro2001}
A. Arbe {\it et~al.}, Macromolecules {\bf 34},  1281  (2001).

\bibitem{AllegraGanazzoli:Review1989}
G. Allegra and F. Ganazzoli,  in {\em Advances in Chemical Physics} (Wiley, New
  York, 1989), Vol.~75, Chap.~Chain configurations and dynamics in the gaussian
  approximation, pp.\ 265--348.

\bibitem{flory2}
P.~J. Flory, {\em Statistical Mechanics of Chain Molecules} (Wiley, New York,
  1969).

\bibitem{WittmerEtal:PRL2004}
J.~P. Wittmer {\it et~al.}, Phys. Rev. Lett. {\bf 93},  147801  (2004).

\bibitem{WittmerEtal:PRE2007}
J.~P. Wittmer {\it et~al.}, Phys. Rev. E {\bf 76},  011803  (2007).

\bibitem{WittmerEtal:EPL2007}
J.~P. Wittmer {\it et~al.}, Europhys. Lett. {\bf 77},  56003  (2007).

\bibitem{BeckrichEtal:Macro2007}
P. Beckrich {\it et~al.}, Macromolecules {\bf 40},  3805  (2007).

\bibitem{SemenovObukhov:JPCM2005}
A.~N. Semenov and S.~P. Obukhov, J. Phys.: Condens. Matter {\bf 17},  S1747
  (2005).

\bibitem{BaschnagelWittmerMeyer:NIC_Review2004}
J. Baschnagel, J.~P. Wittmer, and H. Meyer,  in {\em Computational Soft Matter:
  From Synthetic Polymers to Proteins}, edited by N. Attig, K. Binder, H.
  Grubm\"uller, and K. Kremer (NIC Series, J\"ulich, 2004), Vol.~23, pp.\
  83--140, (available from {\tt http://www.fz-juelich.de/nic-series}).

\bibitem{WittmerEtal:preprint2007}
J.~P. Wittmer {\it et~al.}, Intrachain orientational correlations in dense
  polymer solutions  (preprint).

\bibitem{MeMu01}
H. Meyer and F. M\"uller-Plathe, J. Chem. Phys. {\bf 115},  7807  (2001).

\bibitem{MeMu02}
H. Meyer and F. M\"uller-Plathe, Macromolecules {\bf 35},  1241  (2002).

\bibitem{VettorelMeyer:JCTC2006}
T. Vettorel and H. Meyer, J. Chem. Theory Comput. {\bf 2},  616  (2006).

\bibitem{VettorelEtal:PRE2007}
T. Vettorel, H. Meyer, J. Baschnagel, and M. Fuchs, Phys. Rev. E {\bf 75},
  041801  (2007).

\bibitem{KremerGrest1990}
K. Kremer and G.~S. Grest, J. Chem. Phys. {\bf 92},  5057  (1990).

\bibitem{AuhlEtal:2003}
R. Auhl {\it et~al.}, J. Chem. Phys. {\bf 119},  12718  (2003).

\bibitem{AllenTildesley}
M.~P. Allen and D.~J. Tildesley, {\em Computer Simulation of Liquids}
  (Clarendon Press, Oxford, 1987).

\bibitem{DeutschBinder:JCP1991}
H.-P. Deutsch and K. Binder, J. Chem. Phys. {\bf 94},  2294  (1991).

\bibitem{PaulEtal:JPII1991}
W. Paul {\it et~al.}, J. Phys. II {\bf 1},  37  (1991).

\bibitem{LandauBinder}
D.~P. Landau and K. Binder, {\em A Guide to Monte Carlo Simulations in
  Statistical Physics} (Cambridge University Press, Cambridge, 2000).

\bibitem{Verdier1966}
P.~H. Verdier, J. Chem. Phys. {\bf 45},  2118  (1966).

\bibitem{Edwards_JPhysA1975}
S.~F. Edwards, J. Phys. A: Math. Gen. {\bf 8},  1670  (1975).

\bibitem{SemenovJohner2003}
A.~N. Semenov and A. Johner, Eur.\ Phys.\ J. E {\bf 12},  469  (2003).

\end{thebibliography}

\end{document}